\documentclass[ aps, pra, reprint, superscriptaddress, amsmath, amssymb ]{revtex4-2} \usepackage[utf8]{inputenc}
\usepackage{graphicx}
\usepackage{bm}
\usepackage{comment}
\usepackage[colorlinks=true, allcolors=blue]{hyperref} 
\begin{document} 
\title{Shape Transformations and Rupture of a Vesicle Driven by an Encapsulated Active Particle} 
\author{Gaurav Kumar} 
\email{gaurav@iisermohali.ac.in}
\author{Abhishek Chaudhuri} 
\email{abhishek@iisermohali.ac.in}
\author{Rajeev Kapri} 
\email{rkapri@iisermohali.ac.in} 
\author{Anil Kumar Dasanna} 
\email{adasanna@iisermohali.ac.in} 
\affiliation{Indian Institute of Science Education and Research Mohali, SAS Nagar, Punjab 140306, India}

\large
\begin{abstract}
 Biological membranes are highly deformable structures that undergo shape transformations during numerous cellular processes. These transformations are often driven by active agents, including molecular motors, cytoskeletal machinery, and motile microorganisms, which consume energy and maintain biological systems far from equilibrium. Here, we investigate how a single active particle confined within a lipid vesicle drives membrane deformation. Using a lipid-resolved coarse-grained model, we examine how the activity and shape of an encapsulated self-propelled particle influence vesicle dynamics and morphology. Active particles generate persistent propulsion that exerts mechanical stresses on the enclosing membrane, thereby driving its deformation. Using particles of different shapes but identical surface area, we systematically investigate membrane morphologies as functions of activity and bending rigidity. We quantify membrane deformation through asphericity and excess bending energy relative to the equilibrium spherical state. Both measures exhibit a non-monotonic dependence on activity, peaking at intermediate activity. At sufficiently high activity, membrane rupture occurs before significant large-scale deformation can develop, limiting further remodeling. We further demonstrate that membrane surface tension plays a significant role in regulating activity-induced deformation dynamics. 
\end{abstract}

\maketitle

\section{Introduction}

 Biological membranes are dynamic interfaces that continuously remodel their shape in response to mechanical forces, enabling essential cellular processes such as endocytosis, exocytosis, vesicle trafficking, and cell division \cite{diaconescu2017phosphoinositides,neher2013exocytosis,schiel2011vesicle,gerien2018cytokinesis,shi2015membrane}. These processes involve large-scale changes in membrane curvature and leads to diverse morphologies, including buds, tubes, and vesicles \cite{mcmahon2005membrane,bashkirov2022molecular,day2020biophysical,zimmerberg2006proteins}. Such deformations can arise from localized interactions between membranes and deformable or structured objects: vesicle wrapping is governed by the interplay between membrane bending, adhesion, geometry, membrane tension and deformability \cite{vanderham2025,spanke2022dynamics,ruiz2012mechanisms,xiao2023wrapping}, while coacervates, membrane-binding proteins, and nanoparticles can deform membranes through lipid–polymer, protein–membrane, and particle–membrane interactions \cite{mondal2023,klein2025,bahrami2014wrapping,bahrami2016role}. In particular, interactions with multiple nanoparticles can cooperatively promote membrane tubulation \cite{raatz2014cooperative,weikl2026tubulation}. Together, these studies demonstrate that localized interactions between membranes and external objects can generate a wide range of membrane morphologies even in equilibrium systems.

 In living systems, membrane remodeling is also driven by active agents that consume energy and generate mechanical forces. For example, intracellular vesicles and endosomes are transported along microtubules by molecular motors such as kinesin and dynein. The forces generated by these motors are transmitted to membrane-bound cargo through motor–microtubule interactions and can contribute to membrane deformation, including tubulation and vesicle formation \cite{howard2002mechanics}. Similarly, motile bacteria such as \textit{Listeria monocytogenes} exploit actin polymerization to generate propulsive forces, enabling their movement within host cells and mechanically interacting with and remodeling the surrounding cellular structures \cite{tilney1989actin}. During malaria infection, \textit merozoites actively invade red blood cells through a process involving actomyosin-generated forces and deformation of the erythrocyte membrane. Following invasion, the parasite further remodels the host-cell membrane and its underlying cytoskeleton, altering the mechanical properties of the infected cell~\cite{hillringhaus2019importance,hillringhaus2020stochastic,dasanna2021effect}. In another example of active membrane remodeling, motile \textit{Escherichia coli} encapsulated within giant lipid vesicles exert forces on the inner membrane, producing tubular membrane protrusions \cite{lenagard2022encapsulated}.

 These biological examples illustrate a general principle: energy-consuming active agents can continuously generate mechanical forces that drive membrane deformation and remodeling. Active particles provide an additional mechanism for membrane remodeling by continuously generating mechanical forces through self propulsion. Recent studies have shown that self propelled particles confined within a giant unilamellar vesicles (GUVs) can generate persistent forces that induce large membrane deformations and dynamic shape fluctuations \cite{vutukuri2020active,iyer2023dynamic} by using mesoscopic triangulated membrane model. In these systems, active particles exert persistent forces on the membrane, producing localized deformations whose strength depends on particle activity and membrane particle interactions. Theoretical studies using dynamic traingulated fluid membrane model and active brownian particles have shown that such active forces can induce significant deformation \cite{vutukuri2020active,iyer2023dynamic}.

Despite these developments, the physical principles governing the deformation of membranes by active particles remain poorly understood. Active stresses exerted by active particles on the membrane can drive membrane remodeling and push the system away from equilibrium. The resulting membrane response depends on the balance between this active forcing and the membrane’s intrinsic resistance to deformation, governed by its bending rigidity and tension. The geometry of the active particle further influences how the propulsive force is transmitted to the membrane, providing an additional means of controlling the deformation. As a result, active forcing, membrane mechanics, and particle geometry can act together to produce distinct membrane morphologies. Understanding this interplay is essential for establishing a physical framework for active remodeling of soft membranes.

In this work, we investigate the remodeling of a lipid vesicle induced by a single active particle confined within the vesicle using a coarse-grained membrane model\cite{cooke2005tunable}. The active particle is modeled as a finite-sized cluster of coarse-grained particles, allowing it to directly interact with the membrane while enabling its activity to be systematically varied. We first investigate the effect of particle activity by varying the P\'eclet number and examine the role of membrane bending rigidity in determining the resulting vesicle morphologies. We then consider active particles with different geometries while maintaining comparable surface areas to isolate the influence of particle shape on membrane deformation. The resulting morphologies are characterized using the quantitative measures of vesicle sphericity and membrane bending energy. Finally, we investigate the role of membrane tension in governing membrane remodeling. Our results demonstrate that the resulting membrane morphologies arise from the coupled effects of particle activity, particle geometry, membrane bending rigidity, and membrane tension.\\

 \section{Model}

\begin{figure*}
    \centering
    \includegraphics[width=0.5\textwidth]{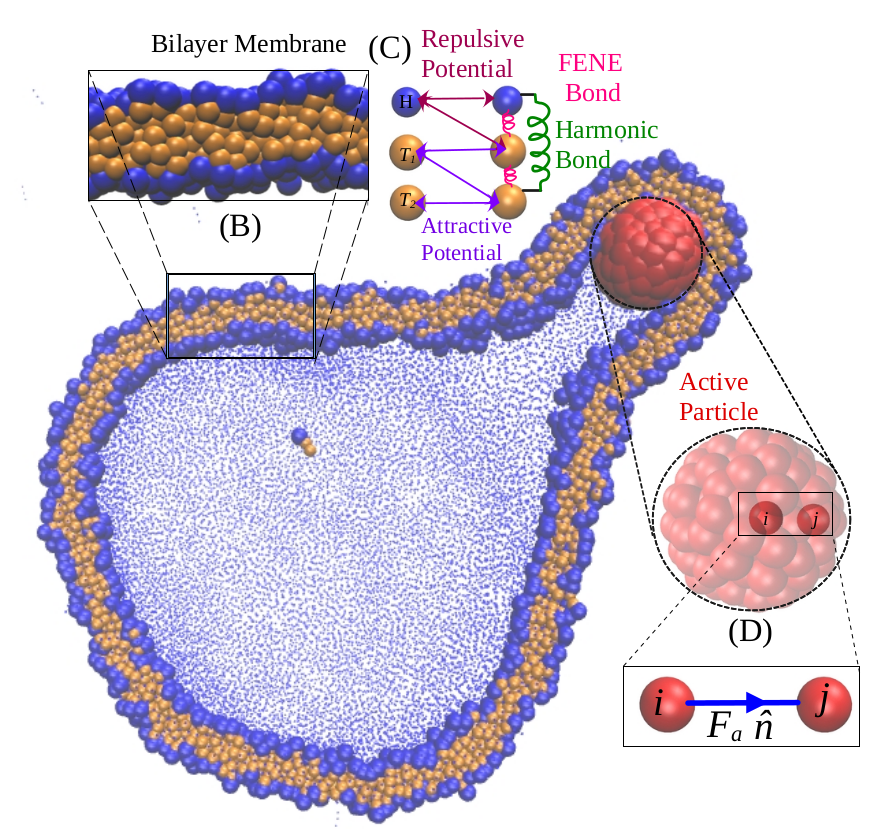}
    \caption{Schematic representation of the coarse-grained simulation model.
    (A) Membrane deformation of a lipid vesicle induced by an encapsulated
    active particle. (B) Solvent-free coarse-grained lipid bilayer membrane
    composed of three-bead lipid molecules. (C) Individual lipid molecule
    showing the bonded and nonbonded interactions responsible for bilayer
    self-assembly. (D) Active particle model composed of rigid surface beads.
    The propulsion direction is defined by the fixed bond between beads
    $i$ and $j$, as indicated by the arrow.}
    \label{fig.1}
\end{figure*}

\subsection*{Vesicle model} 
 We model the vesicle membrane using the coarse-grained three-bead lipid model proposed by Cooke and Deserno \cite{cooke2005tunable,munoz2025tutorial}. In this model, each lipid is explicitly represented by three coarse-grained beads, providing a computationally efficient lipid-level description while capturing the essential physical properties of membranes, including self-assembly, elasticity, and deformation, as illustrated in Fig.~~\ref{fig.1}(A,B). Each lipid molecule consists of one head bead and two tail beads, denoted by $H$, $T1$, and $T2$, respectively, as illustrated in Fig.~~\ref{fig.1}(C). The head bead represents the hydrophilic region of the lipid, whereas the two tail beads represent the hydrophobic chains. The beads within an individual lipid are connected through bonded interactions, while beads belonging to different lipids interact through non-bonded interactions. The combination of these interactions preserves the structure of individual lipids and enables their spontaneous organization into a stable bilayer membrane.\\

The $H-T1$ and $T1-T2$ beads within each lipid are connected by FENE bonds, represented by the pink springs in Fig.~~\ref{fig.1}(C). The FENE potential is given by

\begin{equation}
U_{\mathrm{FENE}}(r) =
-\frac{1}{2} k_{f} R_0^2
\ln \left[1-\left(\frac{r}{R_0}\right)^2\right]\nonumber
\end{equation}

where $k_f$ is the bond stiffness, $r$ is the distance between the bonded beads, and $R_0$ is the maximum allowed bond extension. In our simulations, we use $k_f=30\epsilon/\sigma^2$ and $R_0=1.5\sigma$. The relatively large bond stiffness keeps the beads within each lipid strongly connected, while the finite maximum extension prevents excessive stretching of the lipid molecule. Importantly, the FENE interaction acts only between bonded beads belonging to the same lipid and therefore maintains the molecular integrity of the lipid rather than directly controlling the interactions between different lipid molecules. Here, $\sigma$ and $\epsilon$ represent the characteristic length and energy scales of the model, respectively.

In addition to the FENE bonds, the head bead $H$ and the tail bead $T2$ are connected through a harmonic bond. This bonded interaction constrains the overall geometry of the lipid and helps maintain the preferred orientation of the head and tail regions. The corresponding potential is

\begin{equation}
U_{\text{bend}}(r) =
\frac{1}{2} k_{\text{bend}} (r - 4\sigma)^2\nonumber
\end{equation}

where $k_{\text{bend}}=10\epsilon/\sigma^2$ and $r$ is the distance between the $H$ and $T2$ beads. The preferred separation between these two beads is therefore $4\sigma$. This harmonic constraint, together with the FENE bonds, maintains the elongated geometry of the lipid molecule and prevents excessive conformational distortion during membrane fluctuations and deformation.

The interactions between different lipid molecules are governed by non-bonded potentials. For head-head and head-tail interactions, we employ the purely repulsive Weeks--Chandler--Andersen (WCA) potential

\begin{equation} 
U_{\mathrm{WCA}}(r) = 
\begin{cases} 
4\epsilon \left[ \left(\frac{b}{r}\right)^{12} - \left(\frac{b}{r}\right)^6 \right] + \epsilon, & r < r_{c} \\ 
0, & r \ge r_{c}\nonumber 
\end{cases} 
\end{equation}

where $b=0.95\sigma$ is the effective interaction distance and $r_c=2^{1/6}b$ is the cutoff distance. The WCA interaction accounts for the excluded-volume repulsion between the beads. It prevents different lipid molecules from overlapping and thereby provides the steric interactions necessary for maintaining the local structure of the membrane.

The hydrophobic attraction between lipid tails is introduced through a modified form of the Lennard-Jones potential with a smoothly truncated attractive part,

\begin{equation} 
U_{cos}(r) = 
\begin{cases} 
4\epsilon \left[ \left(\frac{\sigma}{r}\right)^{12} - \left(\frac{\sigma}{r}\right)^6 \right], & r < r_c \\[6pt] -\epsilon \cos^{2} \left( \frac{\pi (r-r_c)}{2w_c} \right), & r_c \le r \le r_c + w_c \\[6pt] 0, & r > r_c + w_c\nonumber 
\end{cases} 
\end{equation}

Here, $r_c=2^{1/6}\sigma$ is the position of the minimum of the Lennard-Jones potential, while $w_c$ determines the range of the attractive interaction. This tail-tail attraction represents the effective hydrophobic interaction between lipid tails and promotes their association and spontaneous formation of a bilayer in the absence of explicit solvent molecules. We express the interaction range in terms of the dimensionless parameter $\tilde{w}_c=w_c/\sigma$ and consider values ranging from $1.5$ to $1.8$. Varying $\tilde{w}_c$ changes the effective mechanical properties of the membrane and provides a convenient means of controlling its bending rigidity; in our simulations, $\tilde{w}_c=1.5$, $1.6$, $1.7$, $1.75$, and $1.8$ correspond to membrane bending rigidities of approximately $8$, $14$, $20$, $25$, and $30,k_BT$, respectively.

\subsection{Active particle dynamics}
The spherical active particle is modeled as a finite-sized cluster of coarse-grained particles with an effective radius of $5\sigma$ and the propulsion direction, $\hat{\mathbf n}$, is defined by a fixed bond between the central particle and a selected particle within the cluster. To investigate the effect of particle shape, we also consider an ellipsoidal active particle with the same surface area as the spherical particle. The ellipsoid has semi major and semi minor axes $a=8\sigma$ and $b=c=3.7\sigma$, respectively. The propulsion direction
of the ellipsoid particle is set along its semi-major axis. Since the cluster is treated as a rigid body, a total active force $F_a\hat{\mathbf n}$ is applied to the cluster, with an equal force $(F_a/N_a)\hat{\mathbf n}$ distributed among its $N_a$ constituent beads. This construction provides a finite-sized active object that can directly interact with the lipid membrane while allowing its activity to be systematically varied. The active particle interacts with the membrane through the Weeks--Chandler--Andersen (WCA) potential described in the previous section, resulting in a repulsive interaction with the membrane lipids. The resulting active force drives the particle out of equilibrium and enables it to exert a persistent mechanical force on the surrounding membrane.

\subsection*{Simulations}
The equation of motion for the active particle is given by

\begin{equation}
m \ddot{\mathbf r}_i =
- \nabla_i U_{\mathrm{tot}}
- \gamma_m \dot{\mathbf r}_i+
F_{a}\,\hat{\mathbf{n}}
+ \sqrt{2 \gamma_m k_B T}\,\boldsymbol{\xi}_i(t)\nonumber
\end{equation}

Here $m$ is the mass of the particle, $\dot{\mathbf r}$ and $\ddot{\mathbf r}$ denote the velocity and acceleration of the particle, respectively. The term $\nabla_i U_{\mathrm{tot}}$ represents the conservative force where $\nabla_i$ is the spatial derivative of the position of the molecule $i$ and  $U_{\mathrm{tot}}$ is the sum of all interaction potentials. $\gamma_m \dot{\mathbf r}_i$ is the viscous term  which  describes viscous damping due to the surrounding implicit solvent and  $\gamma_m$ is the friction coefficient. $F_{a}\,\hat{\mathbf{n}}$ is the active force term which drives are particle far from the equilibrium. $F_{a}$ is the magnitude of the active force and $\hat{\mathbf{n}}$ is the direction of the motion of  particle which is shown in the fig-1(D). The direction of the motion of active particle is fixed by creating a bond between the centre of mass particle $i'$ and a randomly chosen particle $j'$ which is fixed throughout the simulation. The last term represents stochastic thermal forces, where $k_B T$ is the thermal energy and $\boldsymbol{\xi}_i(t)$ is a Gaussian white noise term satisfying $\langle \boldsymbol{\xi}_i(t) \rangle = 0$ and $\langle \boldsymbol{\xi}_i(t)\boldsymbol{\xi}_j(t') \rangle = \delta_{ij}\delta(t-t')$. We use P\'eclet number, a dimensionless quantity that characterizes the relative importance of active motion to thermal diffusion,  to quantify the activity: 

\begin{equation}
Pe = \frac{F_a\, \sigma'}{k_B T}\nonumber
\end{equation}
where $\sigma'=10\sigma$ is active particle size and $k_BT$ is the thermal energy. The membrane remodeling is primarily governed by two key parameters: the membrane bending rigidity, $\kappa$, and the particle activity, characterized by the P\'eclet number, $Pe$. The bending rigidity determines the membrane's resistance to shape deformation, while increasing $Pe$ corresponds to stronger active forcing and, consequently, a greater ability of the particle to deform the membrane. The competition between particle activity and membrane bending rigidity ultimately determines the membrane morphologies observed in our simulations.

We perform molecular dynamics simulations using the LAMMPS simulation package \cite{thompson2022lammps}. In all simulations, the vesicle has a diameter of approximately $60\sigma$, while the active particle has a diameter of $10\sigma$. We systematically vary the membrane bending rigidity, $\kappa$, and the P\'eclet number, $Pe$, to investigate their effects on membrane remodeling and obtain a diverse range of membrane morphologies. The resulting morphological phase diagram is shown in Fig.~2, with the membrane bending rigidity $\kappa$ and the P\'eclet number $Pe$ represented along the horizontal and vertical axes, respectively.

\subsection{Results and Discussion}

\begin{figure*}
    \centering
    \includegraphics[width=0.46\textwidth]{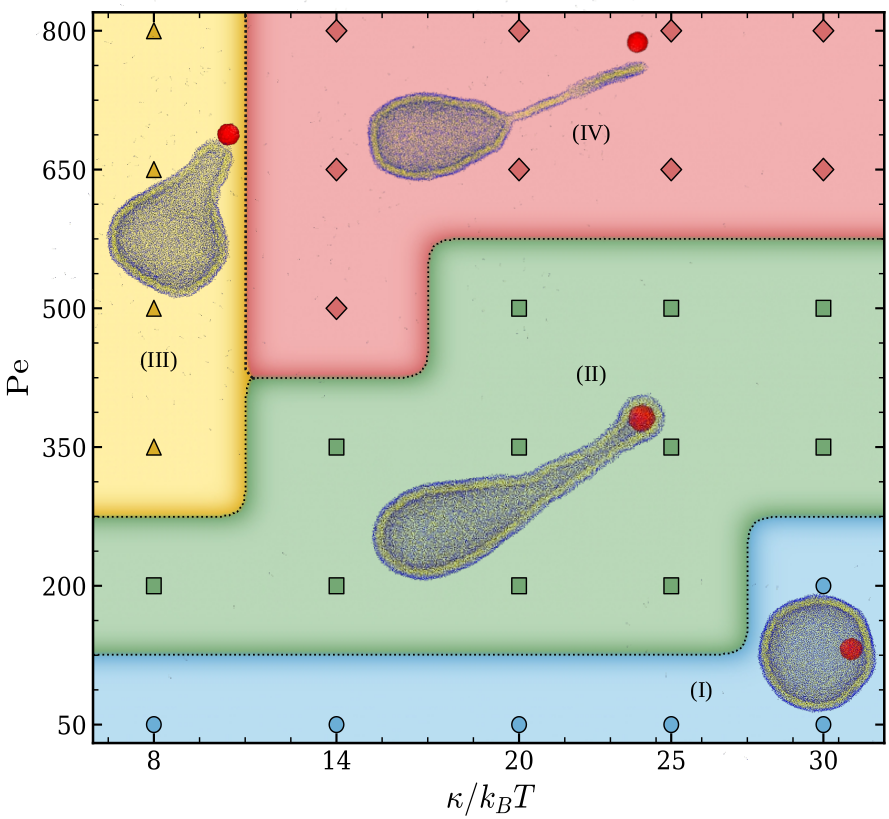}
    \hspace{0.02\textwidth}
    \includegraphics[width=0.50\textwidth]{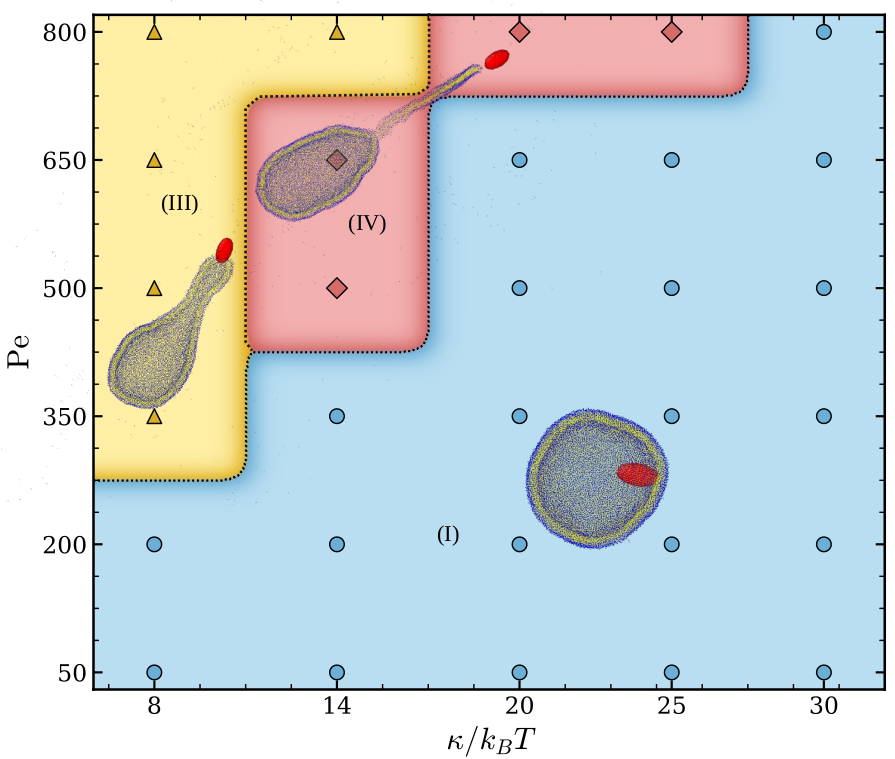}
~~~~~(A)~~~~~~~~~~~~~~~~~~~~~~~~~~~~~~~~~~~~~~~~~~~~~~~~~~~~~~~~~~~~~~~~~~~~~~~~~~~~~~~~~~~~~~~~~~~~~~~~(B)
    \caption{Phase diagrams of vesicle morphologies illustrating membrane
    remodeling of a lipid-bilayer vesicle induced by an encapsulated active
    particle. (A) Results for a spherical active particle with a diameter of
    $10\sigma$. (B) Results for an ellipsoidal active particle with the same
    surface area as the spherical particle, with its propulsion direction
    aligned along the semi-major axis. The membrane bending rigidity,
    $\kappa$, and P\'eclet number, $Pe$, are varied along the horizontal and
    vertical axes, respectively. Different background colors indicate the
    four distinct membrane-remodeling regimes.
   }
    
    \label{fig.2}
\end{figure*}

We investigate the membrane remodeling induced by a single active particle confined within a deformable lipid vesicle. As the active particle continuously propels against the inner membrane, it generates persistent force on the membrane that deform the vesicle and drives the system far from equilibrium. The resulting membrane morphology depends on the competition between the active force generated by the particle and the elastic resistance of the membrane. We identify four distinct morphological regimes, as summarized in Table~\ref{tab:regimes}. By varying the activity of the particle,  we observe a systematic progression from relatively weak membrane deformations to pronounced shape remodeling, including elongated vesicle shapes and tubular protrusions. At sufficiently high activity, the increasing active force produces severe membrane deformation and can eventually lead to membrane rupture, followed by the escape of the active particle from the vesicle.

\subsection*{Geometry-dependent membrane remodeling by active particles}

\begin{table}[htbp]
\centering
\caption{Summary of the four membrane remodeling regimes.}
\label{tab:regimes}

\renewcommand{\arraystretch}{1.5}
\begin{tabular}{|c|c|}
\hline
\textbf{Regime} & \textbf{Membrane Deformation–Escape State} \\
\hline
I & \parbox[t]{0.55\columnwidth}{Nearly spherical membrane; active particle is confined.} \\
\hline
II & \parbox[t]{0.55\columnwidth}{Long tubular deformation; active particle is confined.} \\
\hline
III & \parbox[t]{0.55\columnwidth}{Short tubular deformation; active particle escapes.} \\
\hline
IV & \parbox[t]{0.55\columnwidth}{Long tubular deformation; active particle escapes.} \\
\hline
\end{tabular}
\end{table}

To understand the influence of particle geometry on membrane remodeling, simulations are carried out using two different active particles with identical surface areas but different shapes, namely a spherical particle and an ellipsoidal particle. The membrane morphologies obtained for these two particle geometries are shown in Fig.~\ref{fig.2}(A) and (B), respectively. Although both particles possess the same surface area, their distinct geometries lead to different interactions with the membrane, resulting in qualitatively different deformation pathways and final membrane morphologies. For the spherical active particle Fig.~\ref{fig.2}(A), the vesicle remains nearly spherical at low activity ($Pe=50$), as indicated by the blue color regime in the phase diagram (Movie S1). In this regime, the active force generated by the particle is insufficient to overcome the elastic restoring force of the membrane, and consequently only small membrane fluctuations are observed. As the activity increases to intermediate values of the P\'eclet number($Pe=200-500$), the active particle exerts sufficient force to generate highly curved membrane tubules, represented by the green color regime in the phase diagram (Movie S2). These tubular protrusions emerge from the competition between active force and membrane bending elasticity.

At sufficiently high P\'eclet numbers, the active particle eventually penetrates the membrane and escapes from the vesicle. Interestingly, two distinct escape mechanisms are observed depending on the bending rigidity of membrane. For membranes with relatively low bending rigidity, the membrane is highly deformable, allowing the particle to escape through the formation of a short membrane tubule. These morphologies are represented by the yellow color regime in the phase diagram (Movie S3). In contrast, for membranes with high bending rigidity, the membrane offers greater resistance to bending, and the particle escapes only after generating a much longer membrane tubule. These configurations are represented by the red color regime in the phase diagram (Movie S4). Overall, the phase diagram demonstrates that the interplay between membrane bending rigidity and particle activity governs the transition from a nearly spherical vesicle to tubular morphologies and ultimately to particle escape, thereby controlling the pathways of active membrane remodeling.\\

Fig.~\ref{fig.2}(B) shows the membrane morphologies obtained for an ellipsoidal active particle having the same surface area as the spherical particle. The propulsion direction of the particle is along its semi-major axis. Unlike the spherical particle, the ellipsoidal particle induces significantly weaker membrane deformations over most of the parameter space, and the vesicle largely retains its spherical shape. This behavior arises from the anisotropic geometry of the active particle. When the ellipsoidal active particle interacts with the membrane, the steric repulsion between the particle and the lipid beads reorients the propulsion direction, causing the particle to slide along the inner surface of the vesicle rather than continuously pushing against the membrane at a fixed location. This regime indicated by the blue color in Fig.~\ref{fig.2}(B). Significant membrane remodeling is observed only at high P\'eclet numbers and low bending rigidities, as indicated by the yellow  and red color regimes in Fig.~\ref{fig.2}(B). In these regimes, the active forces overcome the membrane's elastic resistance, leading to the formation of membrane tubules. Compared with the spherical particle, however the ellipsoidal particle requires substantially higher activity to induce similar membrane deformations, demonstrating the important role of particle geometry in regulating active membrane remodeling.\\

The stronger resistance of ellipsoidal particles to membrane deformation highlights the role of particle geometry in regulating active membrane remodeling. This observation is consistent with previous equilibrium studies demonstrating that particle shape, aspect ratio, orientation, and membrane properties influence membrane wrapping and deformation~\cite{dasgupta2013wrapping,bahrami2016role,liu2023wrapping}. In particular, anisotropic particles exhibit distinct wrapping states, with enhanced membrane curvature near regions of larger particle curvature~\cite{dasgupta2013wrapping,dasgupta2014shape}, while the curvature of the vesicle itself can further modulate nanoparticle wrapping~\cite{yu2018nanoparticle}. These geometric effects are also relevant to membrane-mediated interactions, where adsorbed spherical nanoparticles can undergo linear aggregation~\cite{vsaric2012fluid}, potentially promoting membrane tubulation by acting as nucleation sites for tube formation~\cite{vsaric2012mechanism}. More recently, studies of rod-like particles interacting with lipid vesicles have demonstrated that particle curvature, aspect ratio, and orientation influence remodeling pathways ranging from membrane surfing to partial and complete engulfment~\cite{van2024role}. Our results extend these findings by demonstrating that particle anisotropy can suppress active membrane deformation over a broad range of parameters, while sufficiently strong activity can still drive membrane tubulation. The sliding motion induced by steric interactions provides a possible mechanism for this suppression, as it reduces the sustained force exerted on a localized membrane region. Thus, particle geometry not only determines the equilibrium interaction with membranes but also influences how active forces are transmitted to and converted into membrane deformation.

Recent studies of active ellipsoidal particles interacting with giant unilamellar vesicles have shown that propulsion can modify membrane wrapping and induce transitions between partially and completely wrapped states~\cite{tangmongkollert2025dynamics}. Our results extend these findings to an actively propelled particle confined within a vesicle, where particle geometry influences the transmission of active forces to the inner membrane. For ellipsoidal particles, sterically induced reorientation of the propulsion direction promotes sliding along the membrane, reducing sustained force transmission and suppressing deformation. However, sufficiently strong activity can overcome this constraint and induce membrane tubulation. These findings highlight how particle anisotropy regulates active force transmission and membrane remodeling in confined systems.


\subsection*{Quantifying vesicle deformation through sphericity}

\begin{figure*}[tbp]
    \centering
    \includegraphics[width=0.50\textwidth]{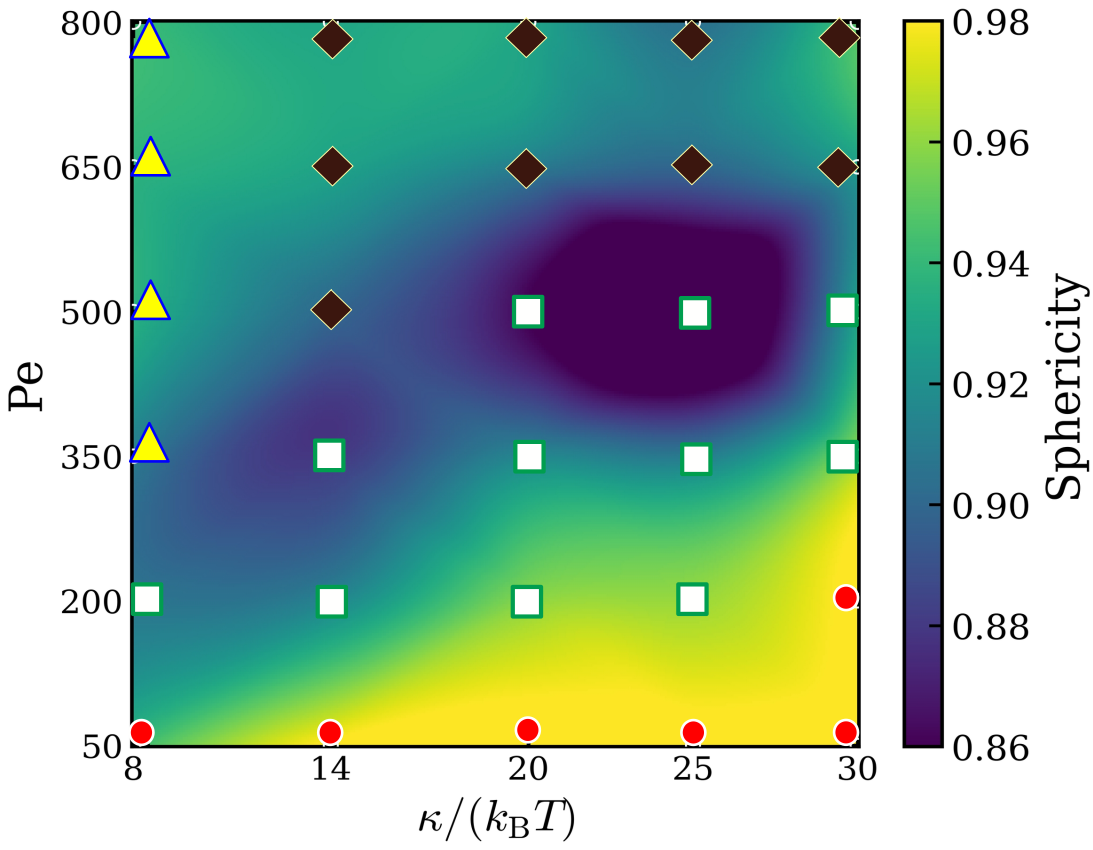}
    \includegraphics[width=0.48\textwidth]{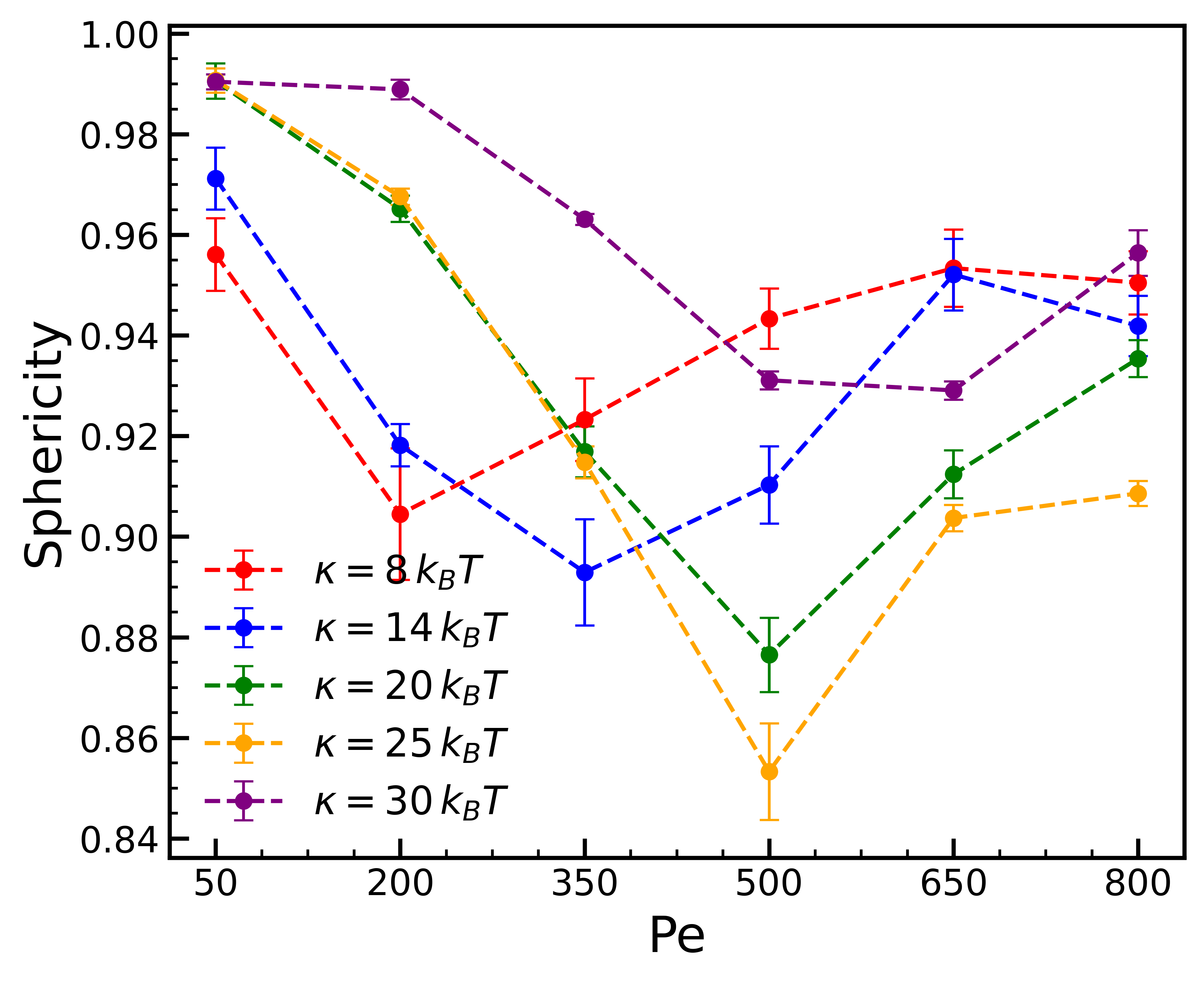}
~~~~~(A)~~~~~~~~~~~~~~~~~~~~~~~~~~~~~~~~~~~~~~~~~~~~~~~~~~~~~~~~~~~~~~~~~~~~~~~~~~~~~~~~~~~~~~~~~~~~~~~~(B)
    \caption{
    (A) Heat map of the vesicle sphericity as a function of the membrane
    bending rigidity, $\kappa$, and P\'eclet number, $Pe$. The color scale
    and numerical values represent the sphericity of the vesicle, which
    quantifies its deviation from an ideal spherical shape. Higher
    sphericity values correspond to nearly spherical vesicles, whereas
    lower values indicate increasingly deformed morphologies.
    (B) Vesicle sphericity as a function of the P\'eclet number, $Pe$, for
    different membrane bending rigidities, $\kappa=8$, $14$, $20$, $25$,
    and $30\,k_BT$. The colored dashed lines represent the variation in
    sphericity with increasing particle activity, and the error bars
    indicate the standard deviation calculated over 100 frames.
    }

    \label{fig.3}
\end{figure*}

We first characterize the vesicle deformation by calculating its sphericity over the range of P\'eclet numbers ($Pe$) and membrane bending rigidities ($\kappa$) considered in our simulations.  Fig.~\ref{fig.3}(A) shows the resulting sphericity heatmap in the $(Pe,\kappa)$ parameter space, where the color scale represents the sphericity of the vesicle. A sphericity of unity corresponds to a perfectly spherical vesicle, while decreasing values indicate increasing deviation from the spherical geometry. At low particle activity, the vesicle remains close to spherical over the entire range of bending rigidities, as indicated by the high sphericity region at low $Pe$. With increasing $Pe$, a region of reduced sphericity develops, particularly for intermediate values of $\kappa$, indicating that stronger active force produces more pronounced membrane deformation. The lowest sphericity region is centered around $Pe\approx500$ and $\kappa\approx20$--$25,k_BT$, where the vesicle undergoes substantial deformation. In contrast, the sphericity remains relatively high at large bending rigidity, particularly near $\kappa=30,k_BT$, demonstrating the ability of a stiffer membrane to resist deformation. The recovery of higher sphericity at large $Pe$ for some values of $\kappa$ also indicates that the response of the vesicle is non monotonic and cannot be described simply by a continuous increase in deformation with particle activity.

To examine these trends more quantitatively, Fig.~\ref{fig.3}(B) shows the sphericity as a function of $Pe$ for individual membrane bending rigidities. At $Pe=50$, the sphericity remains high, with values between approximately $0.96$ and $0.99$ for all values of $\kappa$, indicating that the vesicle is only weakly perturbed by the active particle. As the particle activity increases, the sphericity decreases for most membrane rigidities, although the magnitude of this decrease depends strongly on $\kappa$. For $\kappa=8$$k_BT$, the sphericity decreases from approximately $0.96$ at $Pe=50$ to $0.90$ at $Pe=200$, while for $\kappa=14\,k_BT$ it decreases to approximately $0.89$ at $Pe=350$. A more pronounced deformation is observed for $\kappa=20$ and $25\,k_BT$, for which the sphericity reaches approximately $0.88$ and $0.85$, respectively, at $Pe=500$. In comparison, the stiffest membrane with $\kappa=30\,k_BT$ maintains a relatively high sphericity, decreasing from approximately $0.99$ at $Pe=50$ to $0.93$ at $Pe=500$. Interestingly, for several values of $\kappa$, the sphericity increases again at higher $Pe$, particularly at $Pe=650$ and $800$. This recovery suggests that the vesicle response to increasing activity involves transitions between distinct deformation states rather than a monotonic progression toward increasingly deformed configurations. Overall, Fig.~\ref{fig.3}(A) and Fig.~\ref{fig.3}(B) demonstrate that vesicle morphology is controlled by the coupled effects of particle activity and membrane bending rigidity, with the strongest deformation occurring at intermediate membrane rigidities and sufficiently high particle activity.

\subsection*{Membrane rupture and active particle escape}

\begin{figure*}[tbp]
    \centering
    \includegraphics[width=0.44\textwidth]{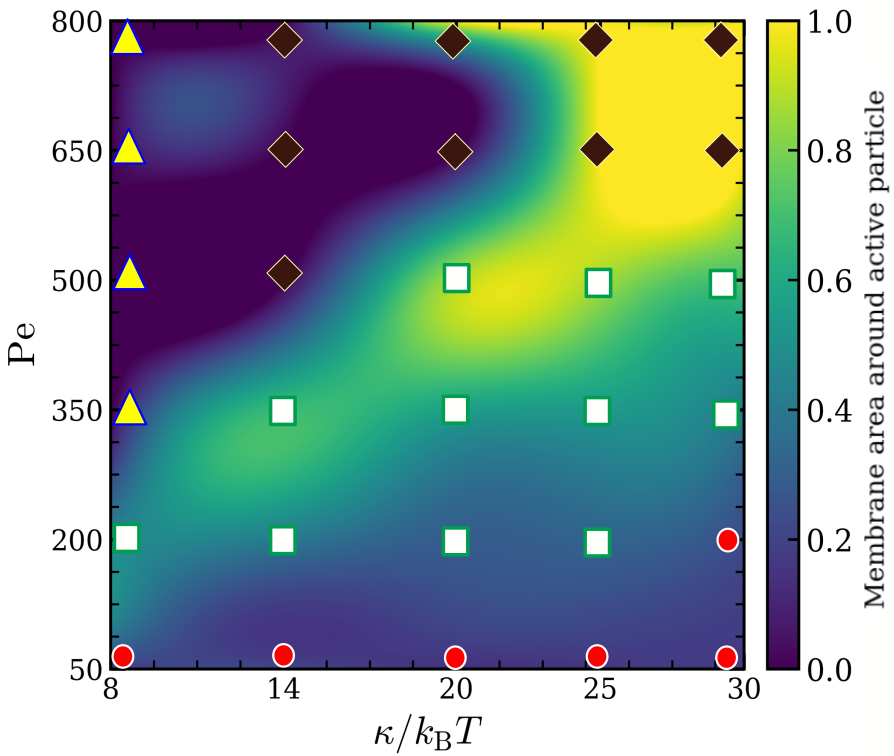}
    \includegraphics[width=0.50\textwidth]{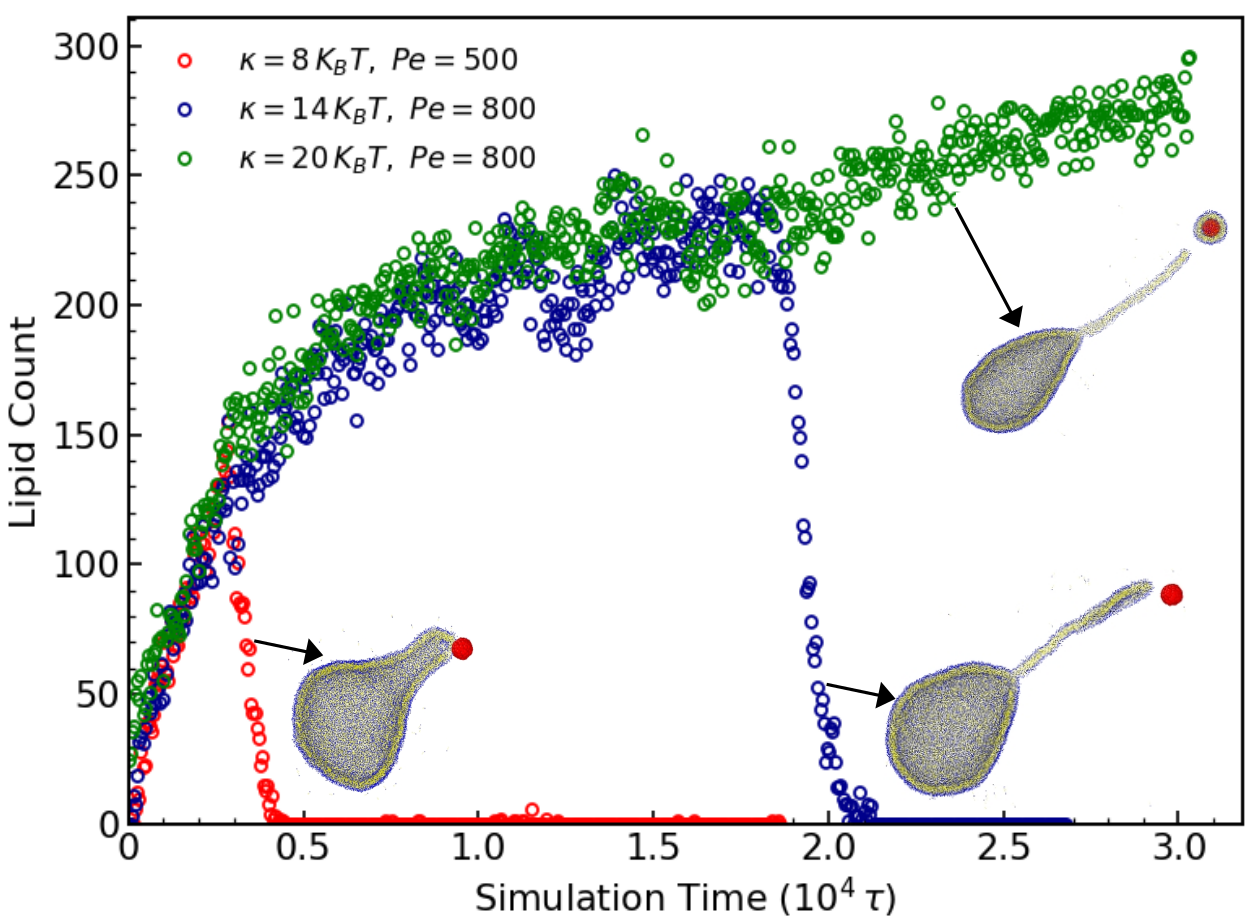}
~~~~~(A)~~~~~~~~~~~~~~~~~~~~~~~~~~~~~~~~~~~~~~~~~~~~~~~~~~~~~~~~~~~~~~~~~~~~~~~~~~~~~~~~~~~~~~~~~~~~~~~~(B)
    \caption{
    (A) Heat map of the membrane area surrounding the active particle as a
    function of the membrane bending rigidity, $\kappa/k_BT$, and P\'eclet
    number, $Pe$. The color scale represents the membrane area surrounding
    the active particle, with larger values indicating stronger membrane
    wrapping around the particle. Symbols indicate the different membrane
    remodeling and particle-escape states observed in the simulations.
    (B) Time evolution of the number of lipid molecules surrounding the
    active particle for representative particle-escape pathways. The blue
    and orange curves correspond to particle escape without membrane
    wrapping, for which the number of surrounding lipid molecules decreases
    to nearly zero after escape. The green curve represents particle escape
    while remaining wrapped by membrane lipids, resulting in a
    membrane-coated active particle after escape.
    }

    \label{fig.4}
\end{figure*}

To further characterize the membrane remodeling process, we investigate how particle activity and membrane bending rigidity influence membrane wrapping and the subsequent escape of the active particle. Fig.~\ref{fig.4}(A) shows the heatmap for membrane area surrounding the active particle as a function of the P\'eclet number ($Pe$) and membrane bending rigidity ($\kappa$). At low particle activity, the membrane area surrounding the active particle remains relatively small, indicating that the particle interacts locally with the membrane without producing substantial membrane deformation. As $Pe$ increases, the membrane area surrounding the particle increases over a broad range of bending rigidities, indicating enhanced membrane deformation and wrapping around the active particle. This effect is particularly pronounced at intermediate and high values of $\kappa$ with higher $Pe$, where the active force is sufficiently strong to generate substantial membrane deformation. The different symbols in Fig.~\ref{fig.4}(A) indicate the distinct membrane remodeling and particle escape states observed in the simulations. 

The corresponding time evolution of the number of lipid molecules surrounding the active particle is shown in Fig.~\ref{fig.4}(B) for representative cases. Two distinct particle escape mechanisms are observed. In the first mechanism, represented by the blue and red curves, the number of surrounding lipid molecules eventually decreases to nearly zero, indicating that the active particle escapes from the vesicle without retaining a membrane envelope. Although the membrane initially deforms around the particle, the surrounding lipids detach during the escape process. In the second mechanism, represented by the green curve, the number of lipid molecules surrounding the active particle continues to increase and remains large throughout the escape process. In this case, the particle escapes while remaining wrapped by a portion of the membrane, resulting in a membrane coated active particle outside the vesicle. This persistent wrapping is closely related to the membrane rigidity and the associated reduction in lipid mobility. The membrane bending rigidity $\kappa$ is controlled by the cutoff parameter $w_c$, with increasing $w_c$ corresponding to larger $\kappa$. The increase in $w_c$ is accompanied by a decrease in the lateral lipid diffusion coefficient(Supplementary: Fig-1), indicating reduced membrane fluidity and a transition toward a more ordered, gel-like state. The reduced lipid mobility favors the persistence of membrane wrapping around the active particle during escape and allowing a portion of the membrane to remain attached to the particle after fission. Thus, the escape pathway is determined by the interplay between active force and membrane properties: sufficiently strong activity can drive membrane fission and particle escape, while the membrane bending rigidity and lipid mobility influence whether the particle escapes alone or remains coated by a portion of the membrane.

While previous studies have demonstrated activity driven membrane deformations and shape transformations in confined vesicles~\cite{vutukuri2020active,iyer2022non,iyer2023dynamic,peterson2021vesicle}, our results reveal that sufficiently strong activity can drive the system beyond membrane remodeling to fission and particle escape. In particular, the two escape pathways observed here--escape without a membrane envelope and escape while retaining a membrane coating--highlight how active force and membrane bending rigidity influence the outcome of membrane rupture. These findings extend previous studies of confined active particles by demonstrating that active membrane remodeling can culminate in distinct particle escape mechanisms.

The transition from deformation to rupture is also consistent with previous studies of driven particle vesicle interactions. Experiments with driven Janus particles have shown that sufficiently strong particle--membrane interactions can split or rupture giant vesicles~\cite{sharma2022driven}. More recently, theoretical work demonstrated that a force-driven inclusion can generate increasing membrane tension and eventually exceed the membrane lysis threshold, resulting in vesicle rupture~\cite{young2026soft}. In our system, increasing activity similarly strengthens the force transmitted to the inner membrane, leading first to enhanced deformation and, at sufficiently high activity, to membrane rupture and particle escape. Thus, our results identify a direct connection between active-force-driven membrane remodeling and particle escape from a confined vesicle.

\begin{figure*}[tbp]
    \centering
    \includegraphics[width=0.47\textwidth]{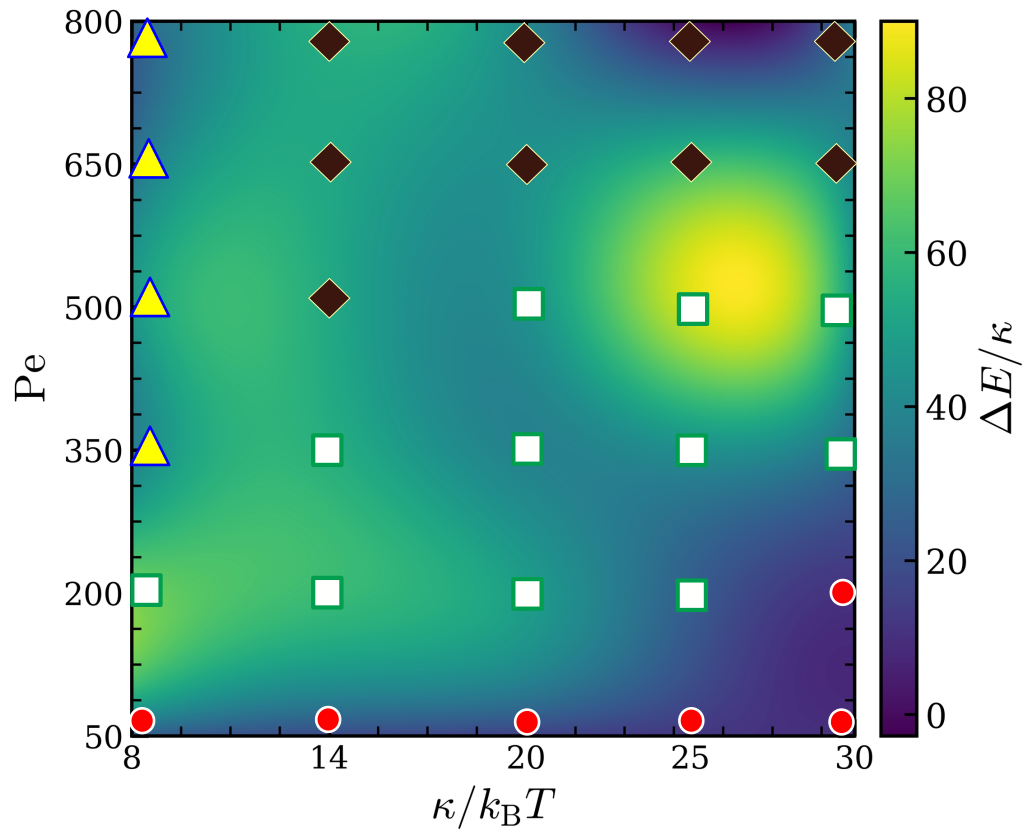}
    \includegraphics[width=0.48\textwidth]{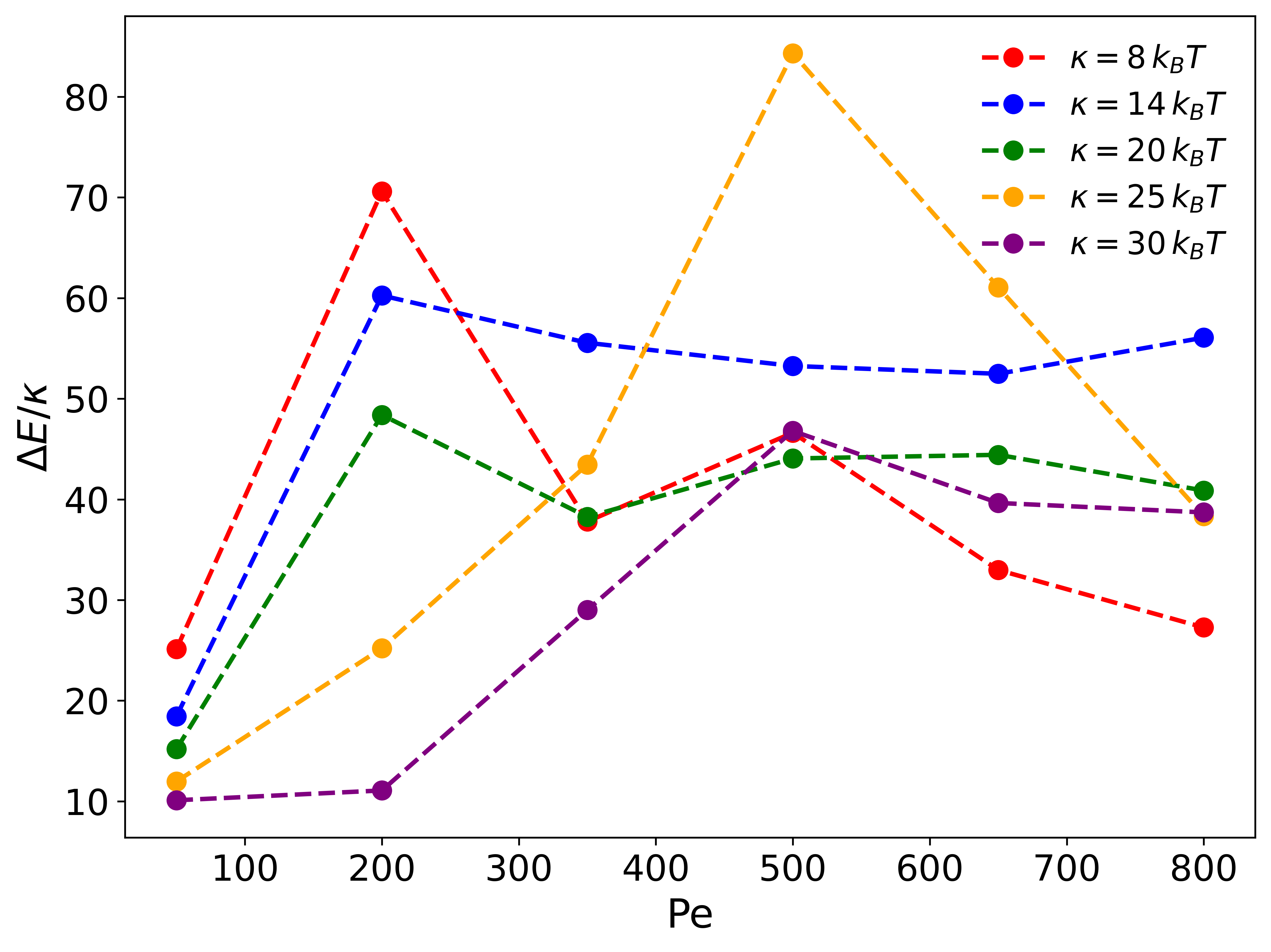}
~~~~~(A)~~~~~~~~~~~~~~~~~~~~~~~~~~~~~~~~~~~~~~~~~~~~~~~~~~~~~~~~~~~~~~~~~~~~~~~~~~~~~~~~~~~~~~~~~~~~~~~~(B)
    \caption{
    (A) Heat map showing the variation of the change in Helfrich bending
    energy, $\Delta E/\kappa$, as a function of the membrane bending rigidity,
    $\kappa/k_BT$, and P\'eclet number, $Pe$. The color scale represents
    the magnitude of the change in bending energy.
    (B) Change in bending energy, $\Delta E/\kappa$, as a function of the P\'eclet
    number, $Pe$, for different membrane bending rigidities,
    $\kappa=8$, $14$, $20$, $25$, and $30\,k_BT$. The nonmonotonic variation
    of $\Delta E/\kappa$ with activity reflects the distinct membrane deformation
    states induced by the active particle.
    }

    \label{fig.5}
\end{figure*}

For further characterization of different membrane morphologies, we quantify the change in the Helfrich bending energy ~\cite{helfrich1973elastic} relative to the corresponding undeformed perfect spherical vesicle. To calculate the bending energy, first we select the head beads of the upper monolayer of the deformed vesicle and reconstruct them as a smooth triangulated surface. In some cases it has been found that when deformation started, leakege of the lipids from vesicle is also started. In those cases the triangulated surface is generated at an early stage of deformation to avoid artifacts of the  missing regions in the membrane surface which will be created due to the significant leakage of the lipids. Fig.~\ref{fig.5}(A) shows the heat map of the change in the Helfrich bending energy, ${\Delta E}$, as a function of the bending rigidity of membrane $(\kappa/k_{B}T)$ and the activity of the particle $(Pe)$. At low bending rigidity, the membrane is more susceptible to  deformation by active particle, resulting in larger fluctuations in the bending energy. As the P\'eclet number $(Pe)$ increases, the active force exerted by the particle becomes stronger, leading to progressively larger membrane deformations and a significant change in the Helfrich bending energy. At higher bending rigidity, the energetic cost of membrane deformation increases, making the vesicle more resistant to the active force, particularly at low P\'eclet numbers. Under these conditions, the vesicle retains its spherical shape, resulting a small variation in the Helfrich bending energy. As the activity strength increases, the active force becomes sufficiently strong to overcome the elastic resistance of membrane, leading to progressively larger deformations. Thus, the heat map clearly demonstrates the competition between membrane elasticity and strength of the active force for the change in the bending energy of the deformed vesicle.
 Fig.~~\ref{fig.5}(B) shows the variation of bending energy with activity strength for different membrane bending rigidities. The bending energy exhibits a non monotonic dependence on $(Pe)$ for all values of $(\kappa)$. At low $(Pe)$, the membrane undergoes relatively weak deformation, while increasing activity enhances membrane deformation and consequently increases  change the bending energy. At higher $(Pe)$, however, the bending energy does not continue to increase monotonically, indicating that the membrane adopts different deformed configurations as the active force becomes stronger. The magnitude and variation of the bending energy also depend on the membrane bending rigidity, reflecting the competition between active force and membrane elasticity.

\begin{figure*}[tbp]
    \centering
    \includegraphics[width=1.0\textwidth]{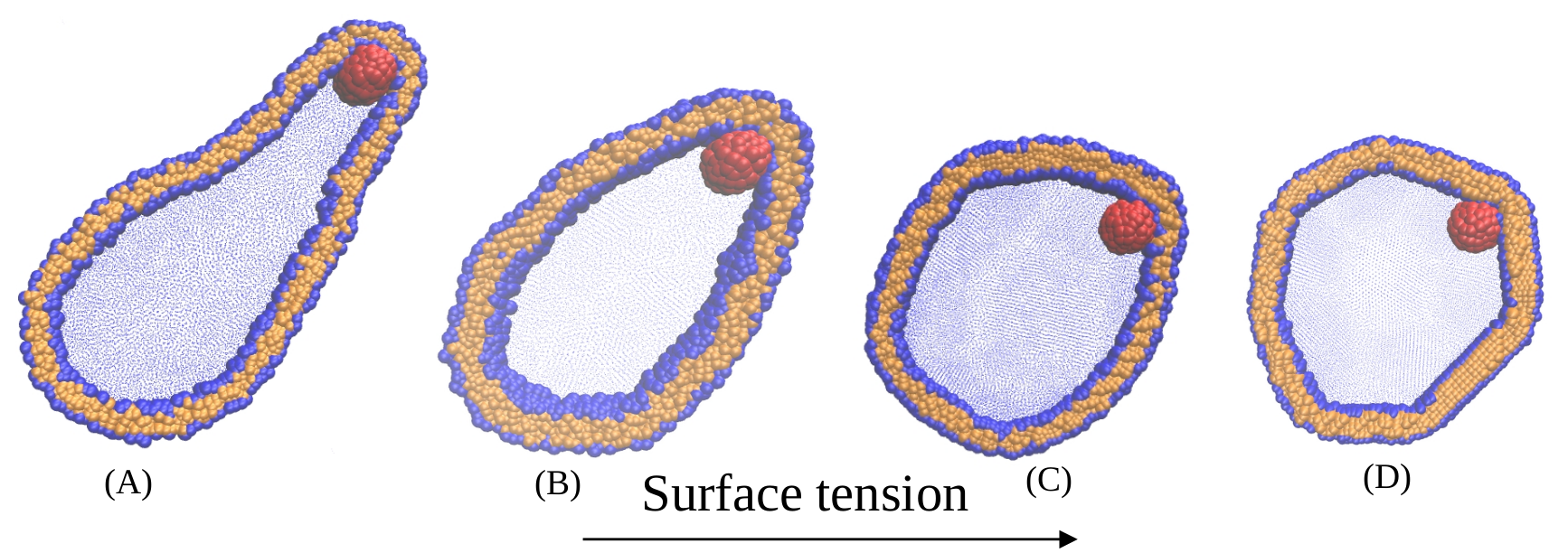}

    \caption{
    Effect of membrane surface tension on vesicle morphology. (A-D)
    Vesicle morphologies obtained by increasing the attractive interaction
    strength between lipid tail beads by $0\%$, $10\%$, $20\%$, and $30\%$,
    respectively, relative to the reference system shown in Fig.~\ref{fig.2}.
    Increasing the tail-tail interaction strength effectively increases the
    membrane surface tension, thereby suppressing membrane deformation and
    promoting a more spherical vesicle morphology. The active particle is
    shown in red, while the lipid head and tail beads are shown in blue and
    yellow, respectively. The membrane bending rigidity is fixed at
    $\kappa=20\,k_BT$, and the P\'eclet number is $Pe=350$.
    }

    \label{fig.6}
\end{figure*}

\subsection*{Effect of membrane surface tension}
To investigate the role of membrane surface tension in the vesicle remodeling by active particle, we systematically increase the attractive interaction strength between the lipid tail beads, thereby effectively increasing the membrane surface tension. The resulting vesicle morphologies are shown in Fig.~~\ref{fig.6} for four different interaction strengths corresponding to 0\%, 10\%, 20\% and 30\% increases relative to the reference system shown in Fig.~~\ref{fig.2}. In all cases, the vesicle initially contains a self-propelled active particle that continuously exerts forces on the membrane, driving it away from its equilibrium spherical shape. The observed membrane morphology is therefore determined by the competition between the active forces generated by the particle and the mechanical resistance of the membrane.\\
The relatively low membrane surface tension allows the active particle to deform the membrane easily, resulting in the formation of a long membrane tube, as shown in Fig.~~\ref{fig.6}(A). The active particle continuously pushes the membrane and deform the vesicle into a membrane tube. As the attractive interaction between the lipid tail beads is increased by 10\% and 20\% (Fig.~~\ref{fig.6}B,C), the effective membrane surface tension increases, making the membrane progressively more resistant to deformation. Consequently, the length of the membrane tube decreases, and the vesicle gradually recovers a more compact shape while still exhibiting localized deformation around the active particle.\\
At the highest interaction strength, corresponding to a 30\% increase in tail-tail attraction (Fig.~~\ref{fig.6}D), the membrane exhibits the highest resistance to deformation. Under these conditions, the active particle is no longer able to generate an extended membrane tube, and the vesicle remains nearly spherical with only a small local depression around the particle. These results demonstrate that increasing the membrane surface tension significantly suppresses membrane deformation by increasing the energetic cost associated with membrane stretching and shape deformation. As a result, higher surface tension inhibits the formation of membrane protrusions and stabilizes the vesicle against large scale morphological changes.


 This behavior is consistent with the general mechanics of membrane tubulation, where the tube radius and the force required for tube formation depend on membrane bending rigidity and surface tension~\cite{helfrich1973elastic,derenyi2002formation,powers2002fluid}. Previous studies have shown that increasing membrane tension suppresses membrane tubulation and shape instabilities~\cite{lipowsky2012spontaneous,kumar2019tubulation,shi2015membrane,kumar2022membrane}. Similar tension-dependent behavior has been reported in active vesicles, where active particles confined within giant vesicles generate pronounced membrane deformations, while increased membrane tension suppresses membrane tether formation~\cite{vutukuri2020active}. Consistent with these observations, our results demonstrate that increasing membrane tension progressively reduces the deformation induced by a single confined active particle, thereby stabilizing the vesicle against large-scale shape changes. These findings highlight the role of membrane tension in regulating the transmission of active forces into membrane deformation, with the deformation response determined by the competition between membrane tension and the magnitude of the transmitted force.

\section{Conclusion}

In this article, we have investigated the deformation of a lipid vesicle induced by a single active particle, which is confined within the vesicle. Here we have used three bead lipid membrane and the activity of the particle is characterized by the P\'eclet number ($Pe$). The active particle continuously exerts a propulsive force on the inner membrane and drives the system far from equilibrium. Our results demonstrate that the different membrane morphologies are governed by the competition between active force and the mechanical resistance of the membrane. By varying the membrane bending rigidity($\kappa$) and P\'eclet number ($Pe$), we observe a transition from nearly spherical vesicles with low activity to strongly deformed morphologies, including membrane tubules and elongated shapes, with higher activity. When the activity is sufficiently high, the active particle ruptures the membrane and escapes from the vesicle.  The membrane response is strongly depends on its bending rigidity. Softer membranes undergo larger deformations with relatively low activity, whereas stiffer membranes oppose deformation and require a higher active force to generate comparable morphological changes. The geometry of the active particle also plays an important role in the vesicle remodeling pathway. A spherical particle can continuously push the membrane and generate pronounced tubular deformations while the ellipsoidal particle tends to slide along the inner wall of the membrane due to its anisotropic geometry. Consequently, the ellipsoidal particle requires activity substantially higher than that of the spherical particle to induce significant membrane deformation. Thus, the shape of the particles provides an additional mechanism for controlling the deformation of the membrane vesicle.

 Furthermore, we have characterized the variation of vesicle sphericity and Helfrich bending energy with activity. Their non-monotonic dependence on the P\'eclet number $Pe$ indicates that membrane deformation does not increase continuously with increasing activity. Instead, the active particle drives the membrane through distinct deformation states, reflecting the competition between active forcing and membrane elasticity. We have also demonstrated that membrane surface tension suppresses activity-induced deformation and stabilizes the vesicle against large-scale shape changes. Together with the dependence on bending rigidity, this highlights the importance of membrane mechanical properties in determining the response to active forcing. At sufficiently high activity, the particle can escape through distinct pathways, either without retaining a membrane envelope or while remaining partially wrapped by the membrane. Previous studies have demonstrated diverse vesicle morphologies induced by multiple active particles using mesoscopic triangulated membrane models \cite{iyer2023dynamic,vutukuri2020active}. In contrast, our study demonstrates pronounced membrane remodeling driven by a single active particle using a coarse-grained lipid membrane model. We identify how particle shape, activity, bending rigidity, and surface tension govern membrane deformation, tubulation, and particle escape, providing insight into the mechanisms underlying activity-driven membrane remodeling.

\section*{Supplementary Movies}

The supplementary movies can be downloaded from 
\href{https://doi.org/10.5281/zenodo.23075415}{here} 
\begin{itemize}
\item \textbf{Supplementary Movie S1:} \underline{Nearly spherical vesicle.}\\
At low value of Peclet number $P_e$, the vesicle remains nearly spherical for all values of the bending rigidity $(\kappa)$. The weak active force is insufficient to produce substantial membrane remodeling. This regime is shown by the blue color in Fig.~2(A).

\item \textbf{Supplementary Movie S2:} \underline{Tubular deformation
	without particle escape.}\\
At intermediate value of Peclet number $P_e$, the vesicle undergoes pronounced deformation and develops a long tubular extension. However, the active particle remains confined within the vesicle and does not escape. This regime is shown by the green color in Fig.~2(A)

\item \textbf{Supplementary Movie S3:} \underline{Particle escape without significant tube formation.}\\
The active particle escapes from the vesicle without generating a long tubular extension with high value of Peclet number ~$P_e$ and low value of bending rigidity $(\kappa)$. The membrane deformation remains localized around the particle, resulting in an absent or very short tube. This regime is shown by the yellow color in Fig.~2(A).

\item \textbf{Supplementary Movie S4:} \underline{Particle escape with a long tubular extension.}\\
At high Peclet numbers, $P_e$, the active particle escapes while strongly deforming the membrane, resulting in a pronounced deformation and an elongated tubular structure. This high activity regime is represented by the red regime in Fig.~2(A).

\item \textbf{Supplementary Movie S5:} \underline{Particle escape with membrane wrapping.}\\
At high values of the Peclet number, $P_e$, and bending rigidity, $\kappa$, the active particle escapes while the membrane lipids wrap around the active particle, resulting in a distinct membrane-wrapped configuration. This behavior can be attributed to the corresponding increase in the cutoff parameter, $w_c$. As $w_c$ increases, the lateral diffusion coefficient of the lipids decreases, indicating a progressive change in the membrane state from a fluid like to a gel like regime. The reduced lipid mobility facilitates membrane wrapping around the active particle during the escape process. This high activity regime is represented by the red region in Fig.~2(A).
\end{itemize}

\subsection*{Lipid Diffusion cofficient}

To characterize the dynamical state of the membrane, we calculated the lateral diffusion coefficient of the lipids for different values of the cutoff parameter $(w_c)$. Fig-1 shows the substantial reduction in lateral lipid mobility, this reduction indicates a progressive change from a fluid like membrane state toward a gel like state with increasing $(w_c)$. This plot provides a useful dynamical measure for identifying the change in membrane state and selecting the appropriate parameter regime for the subsequent simulations. The error bars represent the standard deviation obtained from six independent measurements for each value of $(w_c)$.

\begin{figure}
\centering
\includegraphics[width=80mm]{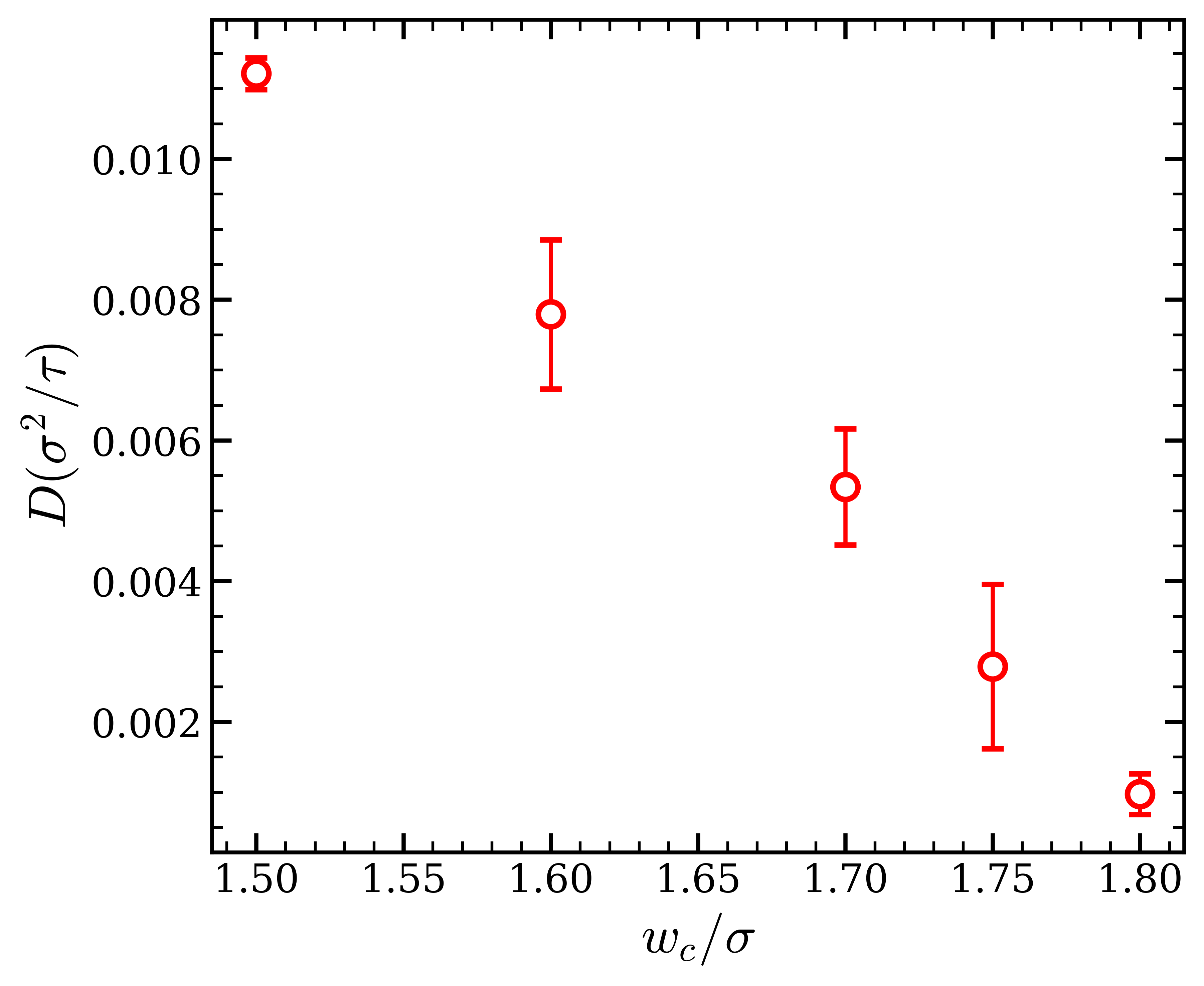}
\caption{Lateral diffusion coefficient $(D)$ as a function of $(w_c/\sigma)$, showing a progressive decrease in lipid mobility and a transition from the fluid-like to gel-like membrane state. }
\label{fig.1}
\end{figure}

\section*{Author Contributions}
Conceptualization: A.C., R.K., A.K.D.; Methodology: A.C., R.K., A.K.D.; Software: G.K., A.K.D.; Investigation: G.K.; Formal analysis: G.K.; Writing – original draft: G.K., A.K.D.; Supervision: R.K., A.K.D.; Interpretation: All authors.

\section*{Declaration of Interests}

The authors declare no competing interests.



\bibliography{refs}

\end{document}